\documentclass[reprint,amsmath,amssymb,aps,twocolumn,superscriptaddress,showpacs]{revtex4-2}
\usepackage{graphicx}
\usepackage{mathrsfs}
\usepackage{bm}
\usepackage{amsmath}
\usepackage{dcolumn}
\usepackage{epstopdf}
\usepackage{dsfont}
\usepackage{amssymb}
\usepackage{tabularx}
\usepackage{array}
\usepackage{float}
\usepackage{color}
\usepackage{epstopdf}
\usepackage{mathrsfs}
\makeatletter
\renewcommand*{\@fnsymbol}[1]{\ensuremath{\ifcase#1\or * \else * \fi}}

\newcommand{\Rmnum}[1]{\expandafter\@slowromancap\romannumeral #1@}
\makeatother
\usepackage[colorlinks=true, letterpaper=true, pdfstartview=FitV, linkcolor=blue, citecolor=blue, urlcolor=blue]{hyperref}
\begin{document}
\title{First-principles study of carrier mobility in MX (M=Sn, Pb; X=P, As) monolayers}

\makeatletter
\renewcommand*{\@fnsymbol}[1]{\ensuremath{\ifcase#1\or *\or \dag\or \dag \or
    \mathsection\or \mathparagraph\or \|\or **\or \dag\dag
    \or \dag\dag \else\@ctrerr\fi}}
\makeatother

\author{Bo Zhang}
\author{Wenhui Wan}
\author{Yong Liu}
\author{Yanfeng Ge}
\email{yfge@ysu.edu.cn}\affiliation{Key Laboratory for Microstructural Material Physics of Hebei Province, School of Science, Yanshan University, Qinhuangdao, 066004, China}


\begin{abstract}
Compounds from groups $\Rmnum{4}$ and $\Rmnum{5}$ have been the focus of recent research due to their impressive physical characteristics and structural stability. In this study, the MX monolayers (M=Sn, Pb; N=P, As) are investigated with first-principles calculations based on Boltzmann transport theory. The results show that SnP, SnAs, and PbAs all exhibit indirect band gaps, whereas PbP is the only semiconductor with a direct band gap. One important finding is that intravalley scattering has a significant impact on electron-phonon coupling. Interestingly, changes in carrier concentration do not affect the electron mobility within these MX monolayers, with SnP exhibiting the highest electron mobility among them. Subsequently, the SnP under a 6\% biaxial strain is further explored and the results demonstrated a considerable increase in electron mobility to 2,511.9 cm$^2V^{-1}s^{-1}$, which is attributable to decreased scattering. This suggests that MX monolayers, especially SnP, are promising options for 2D semiconductor materials in the future.
  \end{abstract}

\maketitle

\section{\label{sec:level1}INTRODUCTION}
Two-dimensional (2D) semiconductor materials have attracted a lot of attention due to their exceptional electronic properties and outstanding structural stability~\cite{1,2}. Among them, graphene stands out for its remarkable carrier mobility, which may reach up to $2\times10^5$ cm$^2V^{-1}s^{-1}$. It also has an excellent elastic modulus and many other extraordinary electronic characteristics that place it at the forefront of research~\cite{3,4}. However, it faces challenges in practical application in nanoelectronic devices due to its small energy band gap at room temperature. Improvements in the field have been facilitated by the exploration and characterization of transition metal disulfides (TMDS) and transition metal trisulfides~\cite{5,6,7,8,9,10,11}. Consequently, group $\Rmnum{4}$ materials, namely silicon and stanine~\cite{12}, as well as group $\Rmnum{5}$ materials, such as phosphorene and arsenene~\cite{13,14,15,16}, have become leading topics in the realm of 2D semiconductor materials research. These discoveries emphasize the ongoing quest for new materials that combine superior electronic properties with appropriate band gaps for useful applications in nanoelectronics.

SnP, a notable metal phosphide material, exhibits impressive physical and chemical properties~\cite{17}. SnP-based nanosheets display outstanding properties such as high switching ratios, great responsivity, and superior photoresponsive qualities. While offering advantages, the synthesis of SnP has presented difficulties in the past. An improved approach has been proposed recently to produce SnP nanorods, surpassing previous methods that relied on hydrothermal synthesis. Researchers have successfully cultivated high-quality 2D SnP nanosheets on liquid metal substrates by using atmospheric pressure chemical vapor deposition~\cite{18}. Subsequently, they managed to control the size of the SnP crystal structure by adjusting the growth temperature and the concentration of the coordination solvent~\cite{19}. The phosphorus/carbon ratio was modified to control nucleation and produce the SnP generated by the carbonization of tin phosphide/carbon nanotubes with carbon nanotubes as initiators. This material demonstrated strong energy storage properties that are superior to those of nanoclusters or nanoparticles, and it was applied to make anode materials for lithium batteries~\cite{20,21}. These processes provide a more reliable and scalable way to manufacture high-quality SnP and related materials. According to theoretical analyses, biaxial strain can efficiently modify SnP’s bandgap while preserving its structural integrity. This makes SnP a viable option for the development of 2D semiconductor devices with good performance.

Carrier mobility is a crucial physical parameter that measures the speed of carrier movement within a material in an electric field, indicating electron and hole migration ~\cite{22}. It is essential for establishing the operational frequency of transistors, the photoconductance gain of photodetectors, and the transmission characteristics of solar cells and other light-emitting devices. In the past, carrier mobility has been calculated by the use of deformation potential theory, which takes into account the electron-phonon interaction caused by longitudinal phonon modes during scattering~\cite{23,24,25,26,27,28}. Nevertheless, density functional perturbation theory offers a more thorough method by considering all electron-phonon couplings~\cite{29,30,31}. In this study, the Boltzmann transport equation is used to calculate mobility. The results show that the electron mobility of the MX (M=Sn, Pb; N=P, As) monolayer ranges from around 312.2 to 528.6 cm$^2V^{-1}s^{-1}$, whereas the hole mobility is between 234.9 to 427.8 cm$^2V^{-1}s^{-1}$. PbAs display intervalley scattering between the $\Gamma$ and M valleys, while intravalley scattering at the $\Gamma$ point primarily contributes to scattering in MX. Moreover, changes in carrier concentration have relatively little effect on carrier mobility in the MX monolayer. Besides, effective modulate carrier mobility may be achieved through strain engineering and carrier concentrations. Notably, SnP reaches 2511.9 cm$^2V^{-1}s^{-1}$ of electron carrier mobility during 6\% biaxial stretching. This improvement is ascribed to changes in the second energy minimum of the conduction band and the energy difference at the conduction band minimum (CBM), which diminish the strength of electron-phonon coupling. With tailored manipulation of carrier mobility, the study offers new insights into the development of high-performance 2D semiconductor devices.

\section{\label{sec:level1} METHODS OF COMPUTATIONAL}
Density functional theory is used in the Quantum-ESPRESSO ground state calculation under the PerdewBurke-Ernzerhof (PBE) exchange-correlation generalization~\cite{32,33}. A conservative pseudopotential was employed and a momentum cutoff of 70 Ry was established. For energy and force, the corresponding convergence criteria are $10^{-4}$ Ry and $10^{-5}$ Ry/a.u. Additionally, a 15 \mbox{\AA} vacuum layer is created to avoid interaction between adjacent atomic layers. Density generalized perturbation theory was applied to determine electronic properties and phonon dispersion on a $12 \times 12 \times 1$ k grid and $4 \times 4 \times 1$ q grid. Subsequently, the maximum localized Wannier functions, encoded in the EPW package~\cite{34,35}, were uniformly interpolated into a dense $120 \times 120 \times 1$ grid to determine the electron-phonon coupling strength and mobility.

The Boltzmann transport equation is applied to calculate the carrier mobility and the final carrier mobility is as follows~\cite{36}. 
\begin{equation}
\mu_{\alpha \beta}=\frac{e^{-}}{n_{\mathrm{e}} \Omega} \sum_{n \in C B} \int \frac{\mathrm{d} k}{\Omega_{\mathrm{BZ}}} \frac{\partial f_{n k}}{\partial \varepsilon_{n k}} v_{n k}, \alpha v_{n k}, \beta \tau_{n k}
\end{equation}
where e, $n_{\mathrm{e}}$ , $\Omega$ , and $\mathrm{k}$ stand for the electron charge, carrier concentration, cell volume, and the electron wave vector integrated into the first Brillouin zone, respectively, while $\alpha$ and $\beta$ denote the  Cartesian coordinates. $\varepsilon_{n k}$ , $f_{n k}$ , and $v_{n k}$ represent electron energy, occupation number, and energy band velocity, respectively. $\tau_{n k}$ is the electronic lifetime, whereas its reciprocal, the scattering rate, is provided by~\cite{37}:
\begin{equation}\begin{aligned}
\frac{1}{\tau_{n k}}= & \frac{2 \pi}{\hbar} \sum_{m v} \frac{\mathrm{d} q}{\Omega_{\mathrm{BZ}}}\left|g_{\mathrm{mnv}}(k, q)\right|^{2} [\left(1-f_{m k+q}+n_{q v}\right) \\
&\times\delta\left(\varepsilon_{n k}-\varepsilon_{m k+q}-\hbar \omega_{q v}\right) \\
&\left.+\left(f_{m k+q}+n_{q v}\right) \delta\left(\varepsilon_{n k}-\varepsilon_{m k+q}+\hbar \omega_{q v}\right)\right]
\end{aligned}\end{equation}
where $\hbar$ is the decaying Planck constant, summed over all phonon indices (v) and energy band indices (m). The symbols q, k, $n_{q v}$, and $\hbar \omega_{q v}$ stand for the phonon vector, electron vector, occupation number, and associated phonon energy, respectively.  The matrix element electron-phonon coupling is represented by $g_{\mathrm{mnv}}(k, q)$. The Dirac delta function ensures energy conservation during the scattering process ~\cite{38}.

\section{\label{sec:level1} RESULTS and DISCUSSION}
\subsection{\label{sec:level2}Electronic band structure and properties.}

\begin{figure}[htbp]
\centerline{\includegraphics[height=0.45\textwidth]{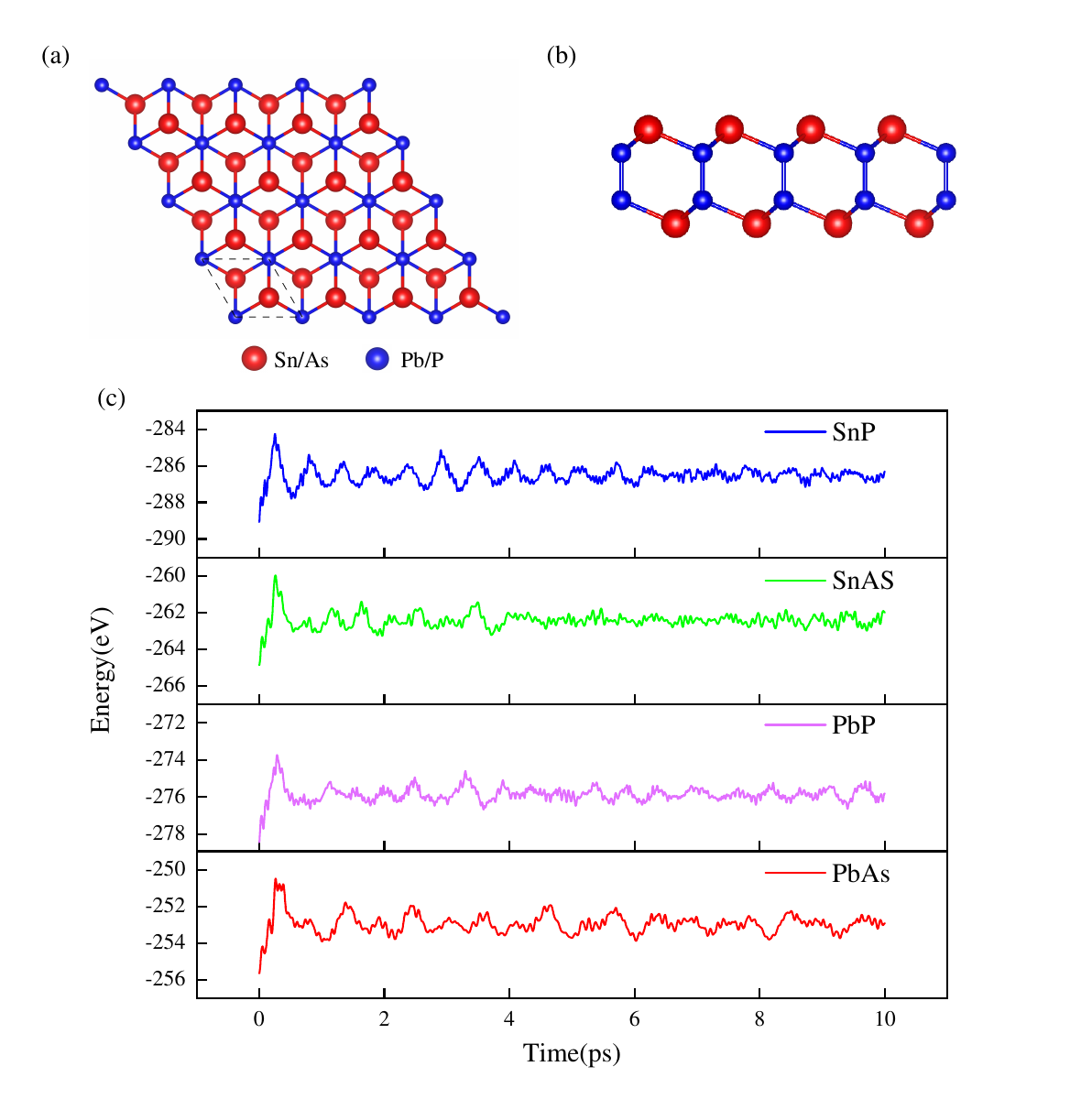}}
\caption{(a) Top and (b) side views of structural MX monolayer. The dashed area denotes a unit cell. The red atoms represent Sn and Pb, and the blue atoms represent P and As. (c) MD simulation of MX at 300 K, the simulation processes 5000 steps at 2 fs per step.
\label{fig:jiegou}}
\end{figure}

The top and side views of Fig.~\ref{fig:jiegou}(a) and (b) exhibit the crystalline structure of the MX (M=Sn, Pb; X=P, As) monolayers, which belong to the P3m1 space group. In these figures, red atoms stand for Sn and Pb, and blue atoms for P and As. As shown by the example of SnP, two Sn atoms live in the outermost layer of P atoms, forming bonds with nearby P atoms. According to the z-direction, this arrangement is Sn-P-P-Sn. The structural optimization led to the determination of the lattice constants of  3.96 \mbox{\AA}, 3.39 \mbox{\AA}, 4.20 \mbox{\AA}, and 4.37 \mbox{\AA} for SnP, SnAs, PbP, and PbAs, respectively. As illustrated in Fig.~\ref{fig:jiegou}(c), molecular dynamics simulations were conducted to evaluate the thermal stability of MX. The findings verify that MX retains an independent and stable monolayer structure at room temperature. Furthermore, Fig.~\ref{fig:ph} displays the phonon dispersion relations, which demonstrate the dynamic stability of MX in the absence of imaginary frequencies.

\begin{figure*}[t!hp]
\centerline{\includegraphics[width=0.75\textwidth]{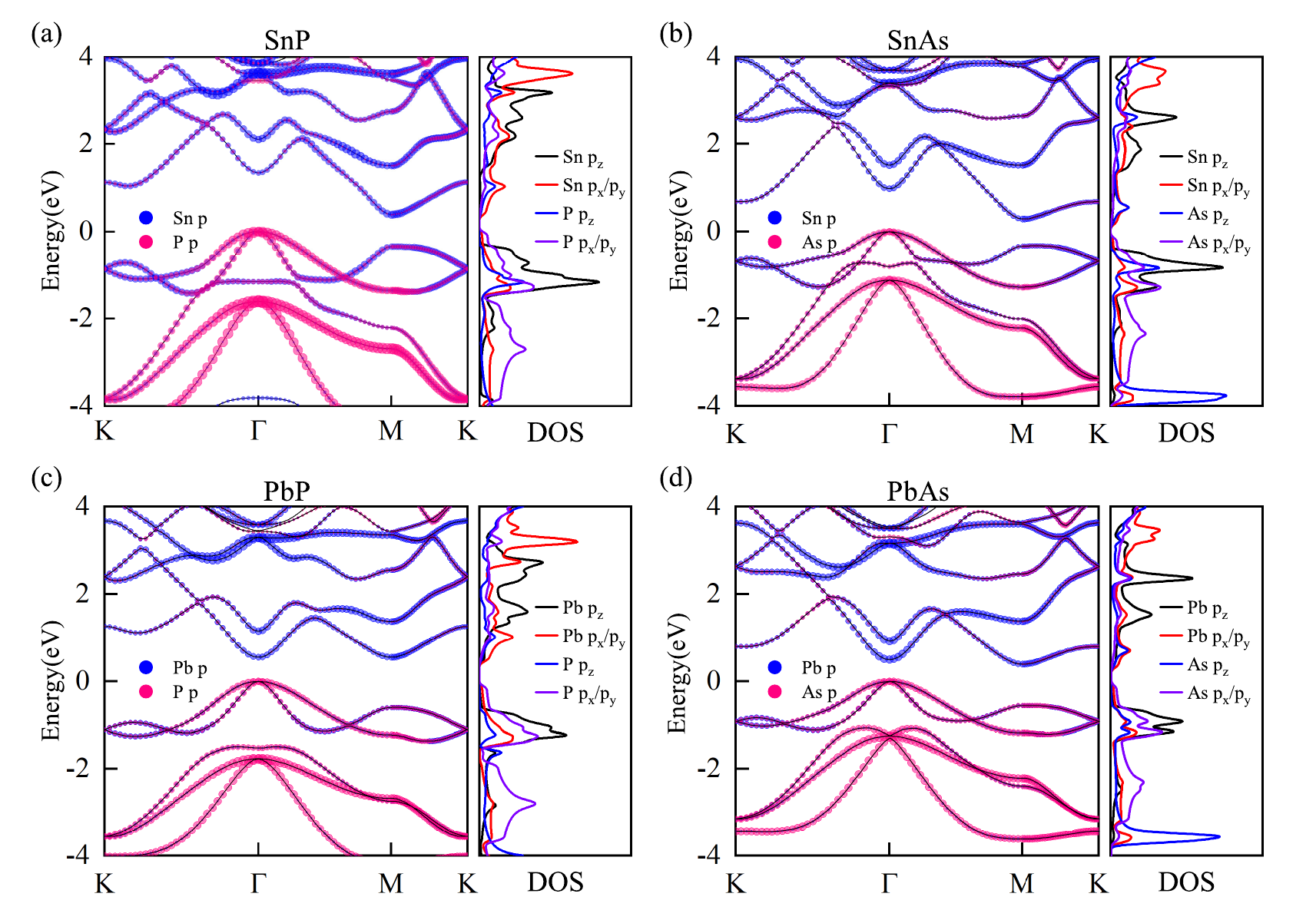}}
\caption{The projected energy band structure and individual atomic contributions in the first Brillouin zone. The bigger the color means more contribution. (a) The projected band structure of Sn and P atoms in monolayer SnP. (b) The projected band structure of Sn and As atoms in monolayer SnAs. (c) The projected band structure of Pb and P atoms in monolayer PbP. (d) The projected band structure of Pb and As atoms in monolayer PbAs.
\label{fig:dos}}
\end{figure*}

As depicted in Fig.~\ref{fig:dos}, SnP, SnAs, and PbAs are recognized as indirect band gap semiconductors. They are distinguished by their valence band maximum (VBM) located at the $\Gamma$ point and their CBM at the M point with band gaps reported at 0.38 eV, 0.24 eV, and 0.39 eV, respectively. On the other hand, PbP is a direct band gap semiconductor that has a band gap of 0.55 eV. Its VBM and CBM are both situated at the $\Gamma$ points. One notable finding across these materials is that the conduction band has a secondary energy minimum. For example, SnP exhibits this secondary minimum at the K’ valley, which is located 66 meV above the M valley and on the line connecting the K and $\Gamma$ points. 
Similarly, the secondary minimum of SnAs is located 40 meV above the M valley at the K valley (directly at the K point). The secondary minimum of PbP’s conduction band is merely 0.1 meV higher in the M valley than in the $\Gamma$ valley.FLastly, the secondary minimum of PbAs at the $\Gamma$ valley is located 10 meV above the M valley. Considering the close spacing between the two energy minima and the large impact of interval scattering on carrier transport, this feature is particularly crucial for our discussion. Atomic orbital contributions were investigated by a projected energy band analysis to corroborate the band gap properties of the MX monolayer. As indicated in Fig.~\ref{fig:dos}, the p$_x$/p$_y$ orbitals of Sn atoms as well as the p$_x$/p$_y$ and p$_z$ orbitals of P atoms contribute mostly to electronic states close to the Fermi energy level in SnP. Moreover, the p$_x$/p$_y$ and p$_z$ orbitals of Sn (Pb) atoms and P (As) atoms are the main contributors to the electronic states at the Fermi energy level in SnAs, PbP, and PbAs.

\begin{figure*}[t!hp]
\centerline{\includegraphics[width=0.75\textwidth]{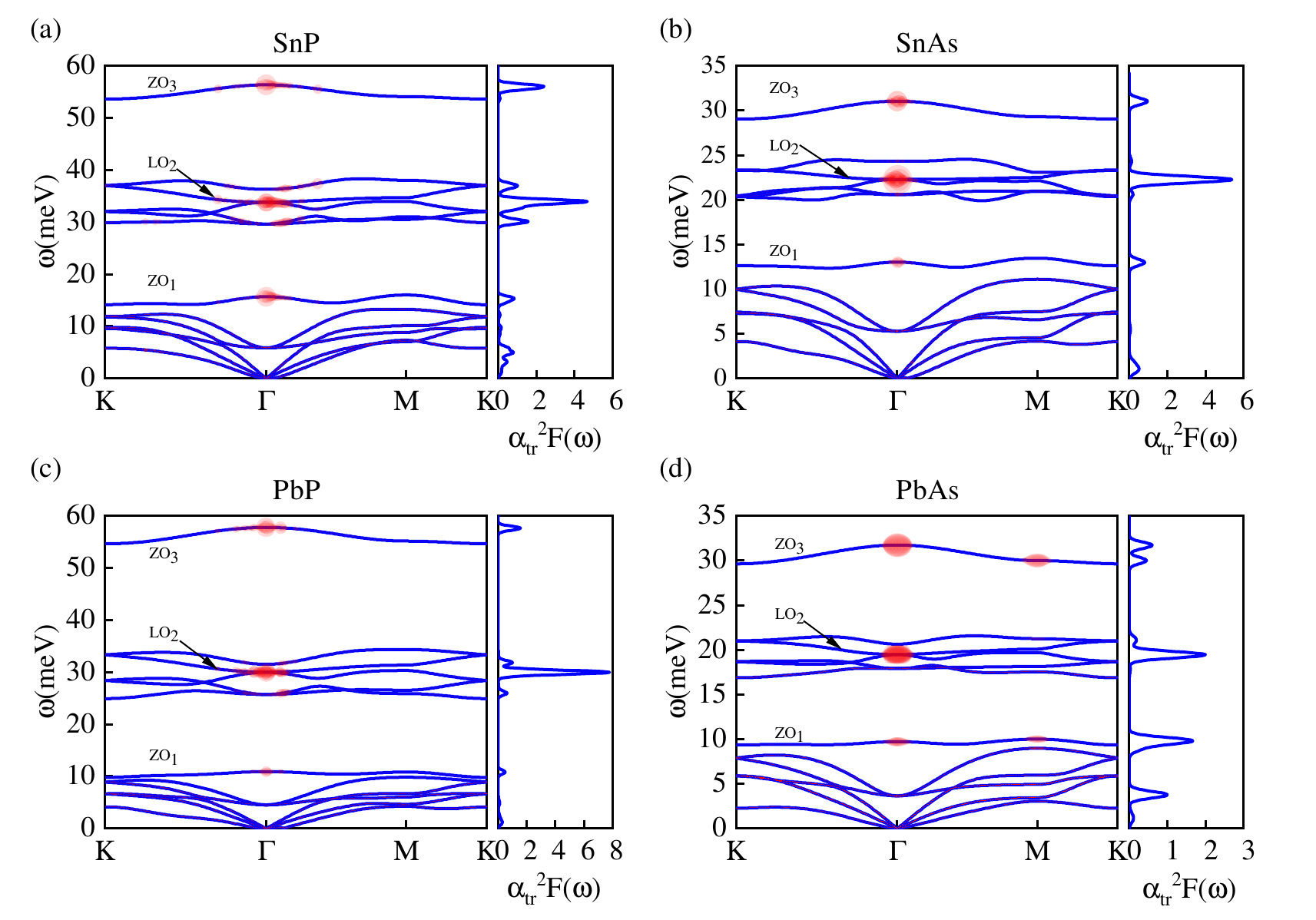}}
\caption{Phonon dispersion of SnP (a), SnAs (b), PbP (c) and PbAs (d) and transport spectral function of electrons $\alpha$$^2_{tr}$F($\omega$) 
\label{fig:ph}}
\end{figure*}

\begin{figure*}[t!hp]
\centerline{\includegraphics[width=0.75\textwidth]{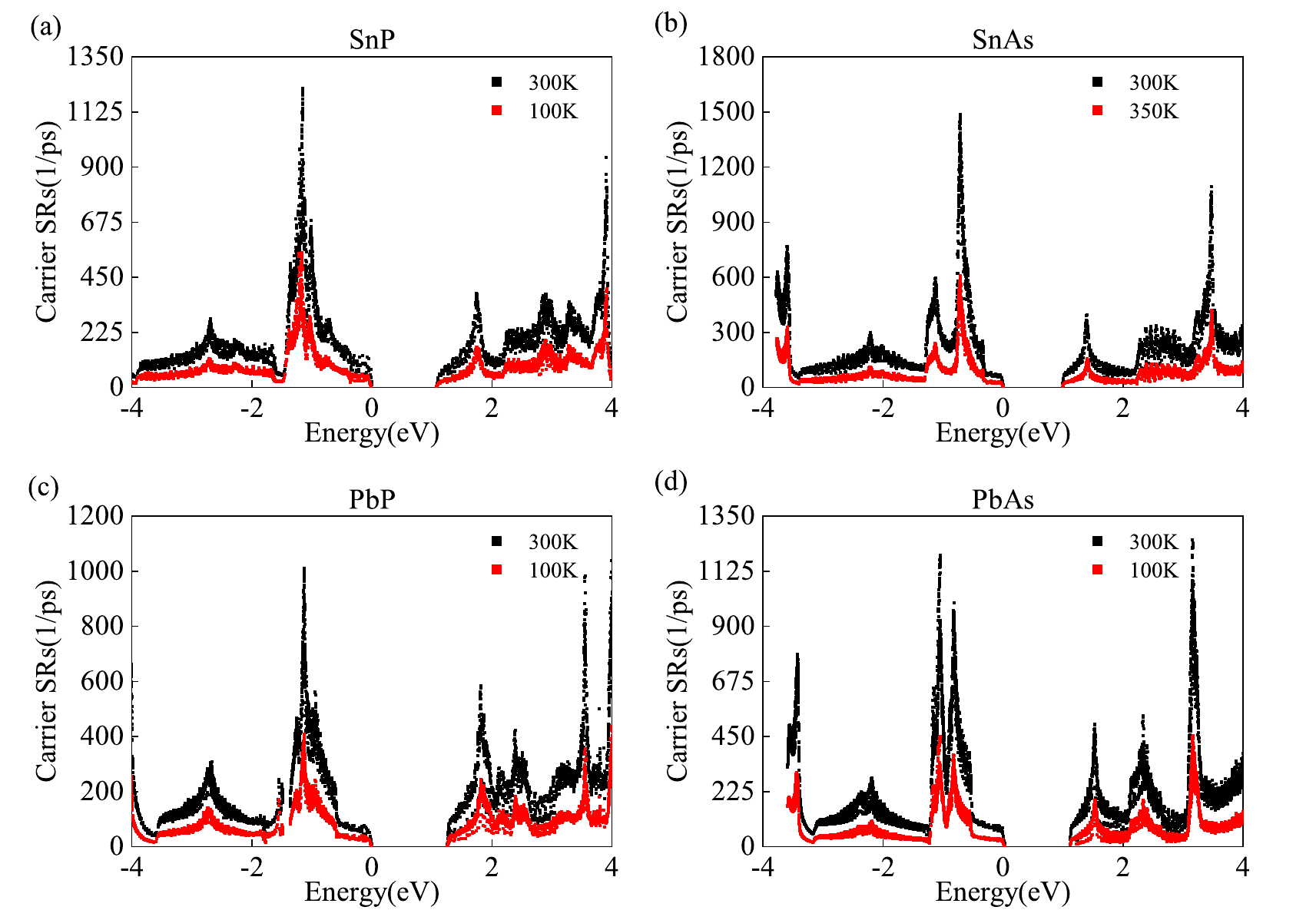}}
\caption{Scattering rates of SnP(a), SnAs(b), PbP(c), PbAs(d) at a carrier concentration of n$_{2D}$=$1 \times 10^{13}$cm$^{-2}$. he colors red and black represent the scattering rates at 100 K and 300 K, respectively.
\label{fig:sanshe}}
\end{figure*}

\subsection{\label{sec:level2}Electron-phonon coupling and carrier mobility.}

\begin{figure*}[htbp]
\centerline{\includegraphics[width=0.75\textwidth]{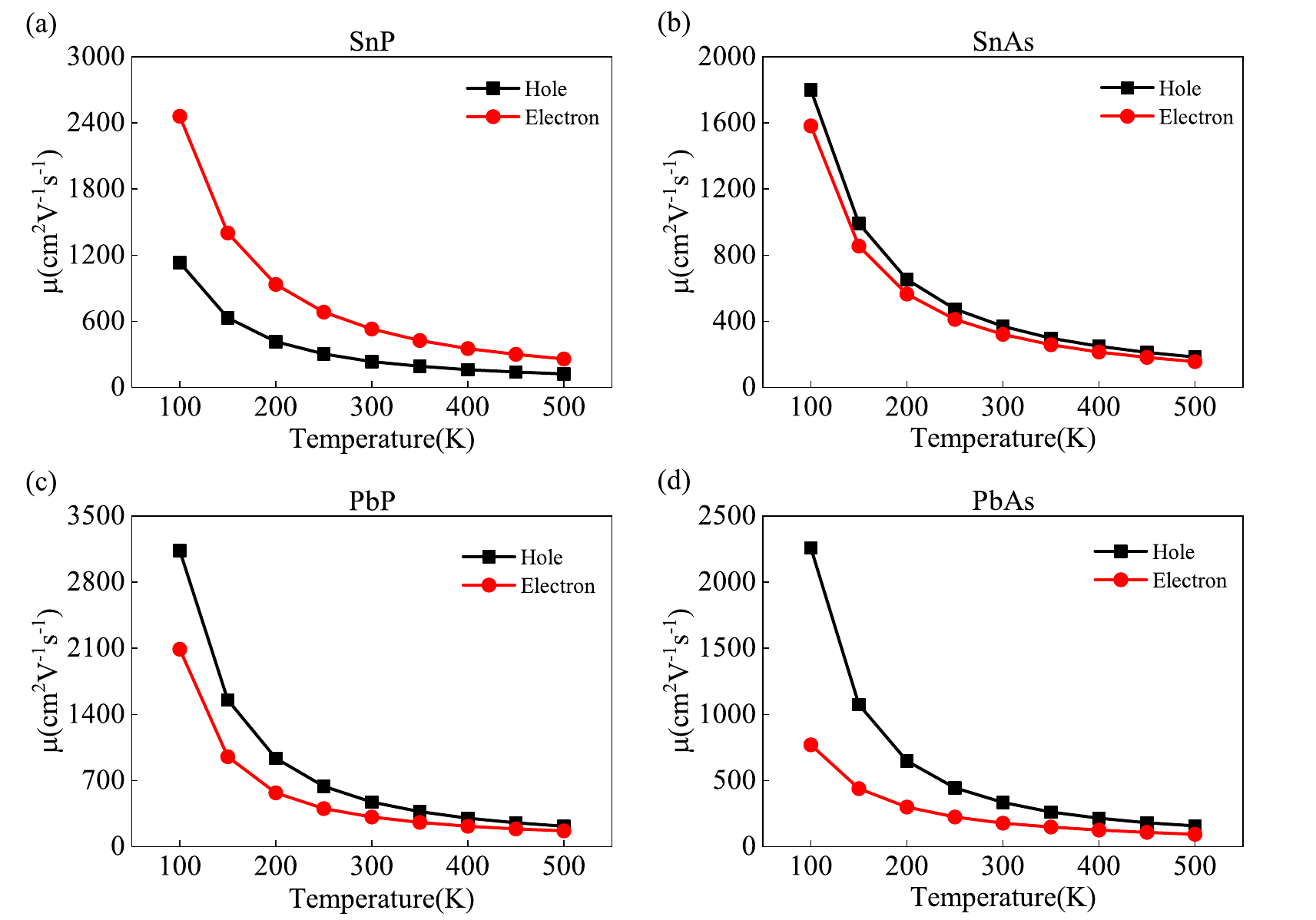}}
\caption{Mobility of electrons and holes at different temperatures for SnP(a), SnAs(b), PbP(c), PbAs(d) when the carrier concentration  n$_{2D}$=$1 \times 10^{13}$cm$^{-2}$. The red circles represent electron mobility, and the black boxes represent hole mobility.
\label{fig:carry}}
\end{figure*}

As shown in Fig.~\ref{fig:ph}, it is essential to understand the phonon dispersion, phonon linewidth, and the $\alpha$$^2_{tr}$F($\omega$) transport spectral function for comprehending the electron transport properties. As the atomic mass of X (P, As) increases, the phonon dispersion curves indicate a decrease in phonon frequencies, particularly for optical phonons. This variation is critical for determining the function of atomic mass in phonon behavior. The degree of electron-phonon coupling is illustrated by the phonon linewidth, which is overlaid on the phonon dispersion diagram. With a distribution that is in line with the $\alpha$$^2_{tr}$F($\omega$) transport spectral function, it highlights key phonon modes that have a major impact on electron-phonon interactions. Specifically, the ZO$_1$, LO$_2$, and ZO$_3$ phonon modes close to the $\Gamma$ point become important in promoting electron-phonon coupling and causing a substantial amount of intravalley scattering. The LO$_2$ optical phonon mode is remarkable due to its dominant influence on electron-phonon coupling among them. Supplementary Fig. S2 displayed the vibrational patterns of optical phonon modes at the $\Gamma$ point. The ZO$_1$ and ZO$_3$ modes include out-of-plane atomic vibrations. As indicated by the peak of the $\alpha$$^2_{tr}$F($\omega$) function, the electron-phonon coupling that controls carrier scattering is primarily induced by the in-plane LO$_2$ mode. The analysis extends beyond the $\Gamma$ point in the case of PbAs, demonstrating the ZO$_1$, LO$_2$, and ZO$_3$ modes at the $\Gamma$ point as well as the important role played by the ZO$_1$ and ZO$_3$ modes at the M point in electron-phonon coupling. This is especially significant to PbAs because of the small energy gap between the CBM and the $\Gamma$ valley. Furthermore, it combines with phonon activity around the M point, enabling significant intervalley scattering from the M valley to the $\Gamma$ valley in the electronic structure.

As the inverse of the carrier lifetime, we first analyzed the total scattering rate of the MX compounds. Fig.~\ref{fig:sanshe} illustrates that the scattering rate of MX at 100 K is substantially lower than at 300 K and increases with rising temperature regimes. In addition, an upward trend in the overall scattering rate is correlated with the atomic mass of X, revealing that optical phonon energy increases as X’s atomic mass grows. Stronger electron-phonon coupling results from this heightened energy, which raises the scattering rate.

At a fixed carrier concentration of n$_{2D}$=$1 \times 10^{13}$cm$^{-2}$, the temperature-dependent changes in electron and hole carrier mobilities for MX compounds are demonstrated in Fig.~\ref{fig:carry}. In particular, SnP has an electron mobility of 528.6 cm$^2V^{-1}s^{-1}$ and hole mobility of 234.9 cm$^2V^{-1}s^{-1}$ at 300 K. The electron and hole mobilities for SnAs are 319.4 and 368.2 cm$^2V^{-1}s^{-1}$, respectively. PbP exhibits electron mobility of 312.2 cm$^2V^{-1}s^{-1}$ and hole mobility of 472.8 cm$^2V^{-1}s^{-1}$. In contrast, PbAs record electron mobility of 178.1 cm$^2V^{-1}s^{-1}$ and hole mobility of 332.5 cm$^2V^{-1}s^{-1}$. As the temperature rises, the data in Fig.~\ref{fig:carry} clearly show that carrier mobility within MX compounds decreases. This trend may be attributed to increased scattering events, which in turn lower carrier mobility. As described in Table S1 of the supplementary material, the effective masses of electron carriers (m$^*_e$) and hole carriers (m$^*_h$) were obtained from the band structure analysis. Among the compounds studied, SnP is noteworthy as an exception because its electron-effective mass is lower than that of hole carriers. On the other hand, the electron-effective mass is greater than the hole carriers for the other three materials. Since carrier effective mass and carrier mobility are inversely related~\cite{39,40}, this implies that hole mobility surpasses electron mobility in all materials studied except SnP.

\subsection{\label{sec:level2} Electronic properties and carrier mobility under strain.}

\begin{figure}[t!hp]
\centerline{\includegraphics[height=0.6\textwidth]{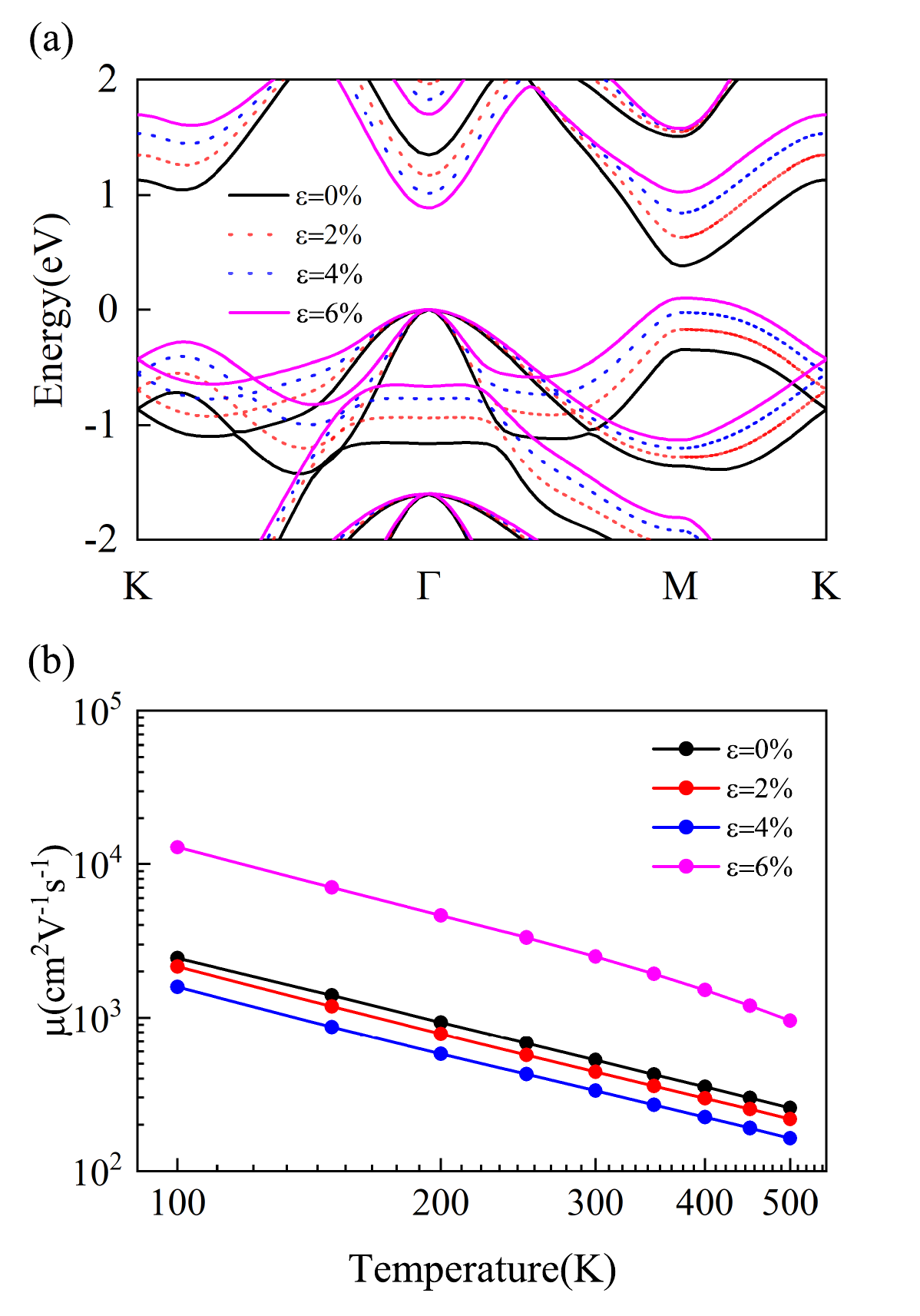}}
\caption{(a) Electron energy band structure of SnP biaxial stretching strain at 300 K and carrier concentration of n$_{2D}$=$1 \times 10^{13} $cm$^{-2}$. (b) The temperature-dependent electron carrier mobility of SnP(n$_{2D}$=$1 \times 10^{13} $cm$^{-2}$) at biaxial tensile strain.
\label{fig:lashen}}
\end{figure}

Previous research has revealed the impact of strain engineering on carrier mobility~\cite{41}. SnP has the greatest electron mobility of all the MX monolayers. Consequently, we performed carrier mobility estimates for SnP at biaxial elongation of 2\%, 4\%, and 6\%. The phonon dispersion and molecular dynamics at these elongations are presented in Figs. S4 and S5 in the supplementary files, which validate that SnP is stable at all three levels of stretching. The electronic energy band structure of SnP during biaxial stretching is displayed in Fig.~\ref{fig:lashen}(a). The VBM and CBM stay at the $\Gamma$ and M points, respectively, at 2\% and 4\% stretching, which correspond to the unstretched state. It is notable that the band gap expands from 0.383 eV to 0.632 eV to 0.845 eV as the strain increases. However, significant shifts take place at 6\% biaxial stretch. 0.784 eV is the band gap caused by the VBM moving from point $\Gamma$ to point M and the CBM shifting from point M to point $\Gamma$. Besides, we found that the interpolation of the second energy minimum of the conduction band with the CBM falls for biaxial stretching of 2\% and 4\%, but increases for 6\%.

As demonstrated in Fig. 6 (b), we investigated the changes in electron carrier mobility under biaxial tensile strain. Interestingly, electron mobility drops at the same level of biaxial stretching as temperature rises. However, electron carrier mobility declines at the same temperature under 2\% and 4\% biaxial stretching, but boots to 2,511.9 cm$^2V^{-1}s^{-1}$ at 6\% biaxial stretching. Two primary factors are responsible for this trend. First, as explained in Table S2 in the supplementary material, the effective carrier mass of electrons increases progressively with 2\% and 4\% biaxial stretching but reduces to 0.22 m$_0$ at 6\% stretching because of changes in the CBM. Second, the difference between the second energy minimum and CBM in the conduction band diminishes at 2\% and 4\% biaxial stretching, which amplifies intervalley scattering. On the other hand, this difference rises at 6\% biaxial stretching, which results in reduced intervalley scattering, lowered electron-phonon coupling strength, and greater carrier mobility.

\section{\label{sec:level1} CONCLUSION}

In conclusion, the MX (M=Sn, Pb; N=P, As) monolayers are investigated using first-principles calculations and the Wannier interpolation technique, offering insights into their electronic structure, electron-phonon coupling strength, and carrier mobility. We discovered that phonon frequency decreases with increasing atomic mass of X, strengthening electron-phonon coupling and influencing carrier mobility. PbAs exhibit intervalley scattering between the $\Gamma$ and M valleys, while intravalley scattering at the $\Gamma$ point primarily contributes to scattering in MX. Intervalley scattering and carrier mobility can be impacted by strain engineering, which alters the energy gap between CBM and the second energy minimum. Electron-phonon coupling is significantly reduced by biaxial stretching of SnP at 6\% strain, leading to an impressive electron carrier mobility of up to 2,511.9 cm$^2V^{-1}s^{-1}$. In addition, our analysis suggests that carrier mobility diminishes as temperature and carrier concentration rise. This work positions MX monolayers, particularly SnP, as promising candidates for next-generation semiconductor applications and lays the way for future research into carrier dynamics and the development of 2D materials.

\begin{acknowledgments}
This work is supported by National Natural Science Foundation of China (11904312, 11904313), Innovation Capability Improvement Project of Hebei province (22567605H), the Hebei Natural Science Foundation (A2022203006), Science Research Project of Hebei Education Department (BJK2022002).
\end{acknowledgments}



\end{document}